\documentclass[preprint,aps,eqsecnum]{revtex4}
\usepackage{graphicx}
\usepackage{latexsym}
\begin{document}
\title{Focusing of branes in warped backgrounds}
\author{Sayan Kar \footnote{Electronic address: {\em sayan@cts.iitkgp.ernet.in}
${}^{}$}}
\affiliation{Department of Physics and Centre for
Theoretical Studies\\ 
Indian Institute of Technology\\ 
Kharagpur 721 302, India}
\vspace{.5in}

\begin{abstract}
Branes are embedded surfaces in a given background (bulk) spacetime.
Assuming a warped bulk, we investigate, in analogy with the case
for geodesics, the notion of {\em focusing} of families of such
embedded, extremal 3--branes in a five dimensional background . 
The essential tool behind our analysis, is the well-known generalised 
Raychaudhuri equations for surface
congruences. In particular, we find explicit solutions of these
equations, which seem to show that families of 3--branes can focus
along lower dimensional submanifolds depending on where 
the initial expansions are specified. We conclude with
comments on the results obtained and possibilities about future
work along similar lines.   
\end{abstract}

\maketitle

\section{Introduction}

The notion of geodesic focusing and the conditions under which it
can occur, follow from the well-known analysis of the Raychaudhuri
equation for the expansion $\theta$ (see \cite{wald}, \cite{poisson}
for further details). Let us briefly recall this analysis, in order
to set the background for further discussion. The Raychaudhuri equation
(for timelike geodesics, assuming shear and rotation to be zero) is given 
as:

\begin{equation}
\frac{d\theta}{d\lambda} + \frac{1}{3} \theta^2 = -R_{ab}\xi^a\xi^b
\end{equation}

where $\xi^a$ is the tangent vector to the geodesic. 
One shows from (1) that if $\theta <0$ for some $\lambda=\lambda_0$ and
if $R_{ab}\xi^a \xi^b \ge 0$, $\theta \rightarrow -\infty$ at a
{\em finite} value of $\lambda$ (focusing). 
One may also convert the above equation
into a second order form by using $\theta=3\frac{F'}{F}$. In the
second order form, one looks for the existence of zeros in $F$ at 
finite $\lambda$, in order to establish the focusing theorem. The criterion 
for the existence of zeros follows from the theory of ordinary differential
equations of the form : $ F''(x)+H(x) F(x)=0$, and is the same as
mentioned at the beginning of this paragraph. Through the
use of the Einstein equation in the convergence condition,
$R_{ab}\xi^a\xi^b\ge 0$ one arrives at the so called strong
energy condition and its variations (eg. weak, null, dominant
energy conditions). Physically, focusing implies the rather
simple fact that in an attractive gravitational field, the
worldlines of test particles tend to converge and ultimately meet. 
Focusing is therefore a pre--condition on the existence of spacetime 
singularities \cite{he}, though it can be 
completely benign, i.e. just a singularity in the congruence, without
there being an actual spacetime singularity at the focal point.  
 
Geodesics are {\em one dimensional} surfaces in a given
background. Their generalisation to higher dimensional embedded
surfaces is provided through the so--called minimal surfaces.
Such surfaces arise as the solutions of the variational problem
for the area action (Nambu--Goto). It is therefore quite likely that
the geodesic deviation (Jacobi) equation and the Raychaudhuri equations 
will also have generalisations for the case of such minimal surfaces.
This was achieved by Capovilla and Guven in \cite{cg}.
Focusing in such a scenario (i.e. a congruence of surfaces
as opposed to a congruence of curves) is the main topic of discussion
in this article. We restrict ourselves to simple examples.
In particular, we assume a five dimensional warped
background in which we have a four dimensional, timelike embedded surface.
This is the currently fashionable braneworld picture \cite{rs}. 
We shall enumerate the various criteria required to obtain results for the
focusing of such a congruence of timelike hypersurfaces in a given warped
background.

In Section II, we review the generalised Raychaudhuri equations for
surface congruences. In Section III, we specialise to 3-branes in a 
warped five dimensional background and obtain solutions of the
generalised equations. Finally in Section IV, we provide some 
remarks and conclude with comments on future directions. 

\section{THE GENERALISED RAYCHAUDHURI (CAPOVILLA--GUVEN) EQUATIONS}

In the case of a family of geodesic curves the quantities that characterise
a flow are the expansion ($\theta$), the rotation ($\omega_{ij}$) and
the shear ($\sigma_{ij}$). The expansion measures the rate of change
of the cross sectional area of a bundle of geodesics. If it goes to
negative infinity within a finite value of the affine parameter, we have
focusing. Shearing corresponds to a change of shape (circle to ellipse,say)
of the cross sectional area and the rotation implies a twist. 
For families of surfaces, these
quantities are replaced by a generalised expansion $\theta_{a}$, a
generalised shear $\Sigma_{aij}$ and a generalised rotation
$\Omega_{aij}$. The index $a$ corresponds to the coordinates of the
embedded surface. Thus, we have an expansion, a rotation and
a shear along each independent direction on the surface. It is obvious that
these quantities are not independent --they are coupled to each other
and that is  what makes their evolution quite difficult to study. 
We shall assume shear and rotation to be zero--this is possible under
certain conditions and we take it that they are satisfied. Thus, we
have only one equation for the generalised expansion given by {\cite{cg}}:

\begin{equation}
\nabla^a\theta_a +\frac{1}{N-D}\theta_a\theta^a + \left (M^2\right )^i_i =0
\end{equation}

where N and D are the dimensions of the background and the embedded
surface and

\begin{equation}
\left (M^2 \right )^i_i = K^{ab i}K_{ab i} + R_{\mu\nu\rho\alpha}
E^{\mu a}E^{\rho}_a n^{\nu i}n^{\alpha}_i
\end{equation}

In the above expression $E^{\mu}_{a}$ constitutes the tangent vector
basis chosen such that $g_{\mu\nu}E^{\mu a}E^{\nu b} = \eta_{ab}$ (a,b run
from 1 to D). $n^{\mu i}$ are the normals, with $g_{\mu\nu}n^{\mu i}n^{\nu j}
=\delta^{ij}$ (i,j run from 1 to $N-D$). Also $g_{\mu\nu}n^{\mu i}E^{\nu a}=0$.
 $K^{ab i}$ are the $N-D$ extrinsic
curvatures (one along each normal direction) 
and $R_{\mu\nu\rho\alpha}$ is the Riemann tensor of the 
background spacetime. Fig. 1 provides a pictorial representation
of the embedding in the $N=3$, $D=2$ case.

One can convert the above equation to a second order form by using
$\theta_a=\partial_a \gamma$ and $\gamma=\left (N-D\right ) ln F$. This
yields, finally :

\begin{equation}
\nabla_a\nabla^a F + \left ( M^2 \right )^i_i F = 0
\end{equation}

which is a variable mass wave equation on the embedded surface. 
Solving this with appropriate initial conditions will therefore
provide the criteria for focusing.

In an earlier paper {\cite{skprd}}, we had arrived at the criterion
for focusing of strings (two dimensional timelike worldsheets, 1--branes)
in an arbitrary background. It turned out that, using a theorem in
the theory of partial differential equations, the existence of
zeros in $F$ seem to be guaranteed if the condition :

\begin{equation}
-{}^2 R +R_{\mu\nu}E^{\mu a}E^{\nu}_{a} >0
\end{equation}

where ${}^2 R$ is the Ricci scalar of the worldsheet.

For the case of 3--branes in a warped background of five dimensions, $N-D=1$
and the quantity $\left (M^2 \right )$ turns out to be :

\begin{equation}
\left (M^2 \right ) = - {}^{4} R +R_{\mu\nu}E^{\mu a}E^{\nu}_a
\end{equation}

If the worldbrane hypersurface is flat then the quantity is further
simplified and we have the equation for the expansion taking the form

\begin{equation}
\nabla_a\nabla^a F + R_{\mu\nu}E^{\mu a}E^{\nu}_a F =0
\end{equation}

For background Einstein spaces with $R_{\mu\nu}= \Lambda g_{\mu\nu}$
it is easy to see that the above equation takes the simple form:

\begin{equation}
\nabla_a\nabla^a F + 4\Lambda F =0
\end{equation}

We shall now try to analyse the abovementioned equations and look for
ways to understand and quantify the notion of focusing of surfaces.

\section{Explicit solutions and the meaning of focusing}

Recall that the background five dimensional line element of a warped
braneworld model {\cite{rs}} is given as :

\begin{equation}
ds^2 = e^{2f(\sigma)} \left (\eta_{ab}dx^a dx^b \right ) + d\sigma^2
\end{equation}

Here, $\sigma$ is usually referred to as the {\em extra} dimension
and the $\sigma=$ constant slice is a four dimensional timelike
hypersurface which is the 3--brane and represents our four dimensional 
world.
 
Assuming the trivial embedding:

\begin{equation}
t=t_1, x= x_1, y=y_1, z=z_1, \sigma = \sigma_0 (constant)
\end{equation}

with $(t_1,x_1,y_1,z_1)$ being the coordinates on the 3--brane, 
we find that the tangents and normals take the simple form:

\begin{eqnarray}
E^{\mu_0} = e^{-2 f(\sigma_0)} \left (1,0,0,0,0 \right ) \\ \nonumber   
E^{\mu_1} = e^{-2 f(\sigma_0)} \left (0,1,0,0,0 \right ) \\ \nonumber   
E^{\mu_2} = e^{-2 f(\sigma_0)} \left (0,0,1,0,0 \right ) \\ \nonumber   
E^{\mu_3} = e^{-2 f(\sigma_0)} \left (0,0,0,1,0 \right ) \\ \nonumber   
n^{\mu} =  \left (0,0,0,0,1 \right ) \\ \nonumber  
\end{eqnarray}

The induced metric on the brane is scaled Minkowski, i.e. 
$\gamma_{ab} = e^{2 f(\sigma_0)}\eta_{ab}$.

\begin{figure}
\includegraphics[width= 9cm,height=5.6cm]{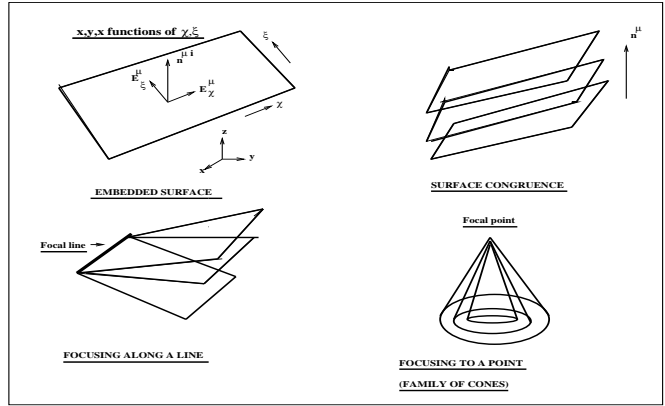}
\caption{The top left figure shows the embedding of surfaces for the case of
a two dimensional one in a 3D background. The other figures demonstrate
a surface congruence, a focal curve and a focal point. The figures are all
qualitative and do not pertain to any of the equations in the article} 
\end{figure}

Using the Ricci tensor components for the background line element, it is
easy to write down the generalised Raychaudhuri (Capovilla--Guven) equations.
We have:

\begin{equation}
e^{-2 f(\sigma_0)} \left ( -\frac{\partial^2}{\partial t^2}+ \nabla^2\right )
F  -4 f''(\sigma_0) F  = 0
\end{equation}

where we have switched back to ($t,x,y,z$) as the coordinates on 
the 3--brane (instead of the ($t_1,x_1,y_1,z_1$) mentioned earlier).
Normally, one does not {\em solve} a Raychaudhuri equation in order
to arrive at a focusing theorem. The usual method is to look for zeros
of F at finite values of the parameters (here $t,x,y,z$) or analyse the
first order equation. However, the
above equation, though a partial differential equation, 
is simple enough and we may solve it directly to get
further insight.
But, to solve the above equation, we do need boundary conditions. The 
central question is, where do we impose those conditions ? Let us first
try a simple solution of the form :

\begin{equation}
F= A \cos \left ( {\vec k}.{\vec x}-\omega t + \phi_0 \right )
\end{equation}

where $A$ and $\phi_0$ are constants. It is easy to see that $k$ and
$\omega$ will be related via a `dispersion relation' of the form :

\begin{equation}
\omega^2= k^2 + 4f''(\sigma_0)e^{2 f(\sigma_0)}
\end{equation}

Here, it is worth noting that the non--trivial curvature of
the five dimensional background makes it appearance only through the above
$\omega-k$ relation. 
The geometric stability criteria discussed in \cite{pk06}, in addition, 
implies $f''(\sigma_0) >0$.
The above solution for $F$ would lead to the following expressions for the
four expansions $\theta_a$. We have:

\begin{eqnarray}
\theta_t = \omega \tan \left ({\vec k}.{\vec x}-\omega t + \phi_0\right ) \\
\theta_{(x,y,z)} = -k_{(x,y,z)} \tan \left ({\vec k}.{\vec x}-\omega t + \phi_0
 \right )
\end{eqnarray}

One immediately notices a couple of important facts from the above. 

(i) Assuming an initial condition defined on the 2--brane given by the 
equation ${\vec k}.{\vec x}-\omega t + \phi_0 = \phi_{(init)}$ one can see that
there is a possibility of a divergence to negative infinity 
(starting with a finite  negative value of $\theta_t$) 
in $\theta_t$ as one approaches the 2--brane
surface ${\vec k}.{\vec x}-\omega t + \phi_0 = \frac{\pi}{2}$. 
$\theta_{(x,y,z)}$ is however is positive initially and diverges
to positive infinite at ${\vec k}.{\vec x}-\omega t + \phi_0 = \frac{\pi}{2}$.  

(ii) Both $\theta_t$ and $\theta_{(x,y,z)}$ cannot simultaneously go to
negative infinity because if one assumes a initially negative expansion
for one, the other has to be positive and vice versa.

The above discussion can also be carried out assuming the solution for
F as $F=A \cos \left ({\vec k}.{\vec x}+\omega t +\phi_0\right  )$. This
will correspond to a time reversed situation. The general results are
similar with reversal of signs at appropriate places. Alternatively, assuming
a sinusoidal form or a linear combination of sines and cosines does not
yield anything newer. In either case,
the point to note is that if we provide {\em initial conditions on 
a 2--brane surface} then the focusing is on the
2--brane. 

If focusing is defined to be the approach to negative infinity
(at a finite value of the coordinates that define the surface) of {\em all}
the expansion vector components $\theta_a$, then we {\em do not} have 
focusing in this case.
In other words, giving initial values on a 2--brane does not yield a
focusing in the sense defined above. However, one can also argue that
$\theta_a$ being a vector field on the embedded surface (here, the 3--brane)
the divergence of any one component would correspond to a singularity in the
surface congruence. 
Thus, one may extend the usual notion of a focal point 
to include {\em focal curves} and {\em focal surfaces}.
These features are illustrated in Figure 1 for the $N=3$, $D=2$
case.  

Another worthwhile question to ask is-- can we have solutions such that 
(i) none of the expansions diverge anywhere (ii) all the expansions diverge 
(to negative infinity) somewhere?
The answer to (i) is straightforward. Assume a solution of the
form :

\begin{equation}
F= A \cosh \left ({\vec k}.{\vec x} -\omega t + \phi_0\right )
\end{equation}

This, when substituted in the equation for F, will give a `dispersion
relation' of the form: 
 
\begin{equation}
\omega^2= -k^2 - 4f''(\sigma_0)e^{2 f(\sigma_0)}
\end{equation}

Here, we require $f''(\sigma_0) <0$ (which goes against the
criterion of geometric stability discussed in \cite{pk06}) 
in order to have anything meaningful.
Furthermore, we need $4 \vert f''(\sigma_0)\vert > k^2$. With these
assumptions (if at all permissible) one can show that the expansions 
are all proportional to 
$\tanh \left ({\vec k}.{\vec x} -\omega t + \phi_0\right )$ and, therefore,
they are all finite everywhere. 

The answer to the second issue (ii) is somewhat more involved. To see this,
let us assume a solution in the form :

\begin{equation}
F= T(t) X ({\vec x})
\end{equation}

This will yield ordinary differential equations of the form:

\begin{eqnarray}
\ddot T + \left (\omega^2 + 4 f''(\sigma_0) e^{2f(\sigma_0)} \right ) T=0 
\\ 
\left ( \nabla^2 + \omega^2 \right )F =0 \\
\end{eqnarray}

If $\omega^2=k^2>0$ and $f''(\sigma_0) >0$, then both the solutions can be
oscillatory and, typically, we have:

\begin{eqnarray}
\theta_t = -\omega_0\tan \left [ \omega_0 t +\phi_0\right ]
\\
\theta_{(x,y,z)} = -k_{(x,y,z)}\tan \left [ {\vec k}.{\vec x} +\phi_1\right ]\\
\omega_0 = \sqrt{4f''(\sigma_0)e^{2f(\sigma_0)}+\omega^2}
\end{eqnarray}    

From the above, it is apparent that if we impose an initially negative 
expansion at some $t=t_0$ and on ${\vec k} .{\vec x} +\phi_1=d$ we can have all expansions
going to negative infinity at finite values of $t$ and $x,y,z$. The focal
surface, in this case would be spacelike and two dimensional.

One can go further and assume solutions of the form :

\begin{equation}
F= T(t) X(x) Y(y,z)
\end{equation}

or, 

\begin{equation}
F=T(t) X(x) Y(y) Z(z)
\end{equation}

In the first of these, one finds a focal curve (one dimensional) and the 
final one we have a focal point (zero dimensional). Let us evaluate the
expansions for the latter. These turn out to be:

\begin{equation}
\theta_x = -\sqrt{\omega^2+\omega_1^2} \tan \left (\sqrt{\omega^2+\omega_1^2} x
\right ) 
\end{equation}
\begin{equation}
\theta_y = -\omega_2 \tan \omega_2 y 
\end{equation}
\begin{equation}
\theta_z = \sqrt{\omega_1^2+\omega_2^2} \tanh \left (\sqrt{\omega_1^2+\omega_2^2} z \right ) 
\end{equation}
\begin{equation}
\theta_t = -\sqrt{\omega^2+4 f''(\sigma_0) e^{2 f(\sigma_0)}} 
\tan \left (\sqrt{\omega^2+4 f''(\sigma_0) e^{2 f(\sigma_0)}} t \right )
\end{equation}

where $\omega_1^2$ and $\omega^2$ are appropriate separation constants.
It is easy to see that in the above expressions for $\theta_a$ one {\em cannot}
have all of them going to negative infinity even if we asuume all of them
to be initially negative. In fact, one of the components (here $\theta_z$)
never diverges. Thus, providing a initially negative expansion vector field
(at a given point) can lead to a focusing at a point though all the expansion
coefficients will not diverge there. 

Thus, the important lesson we have learnt is that the notion of focusing
of surfaces is quite different (as it should be!) from the case of
geodesic curves. The main difference arises because of the fact that 
there are several coordinates (parameters) that define a surface as
opposed to the single parameter that is required to define a curve.
The expansion, is thus a vector field defined on the surface and 
the divergence of its components at some value or set of values
would indicate focusing. The 
notion of focusing along a special submanifold (as illustrated
above) depends on where (i.e. on which submanifold) we impose the initial
condition. We have also noted that all components of the
vector field may not diverge to negative infinity.  

In the above example, we assumed that the induced metric on the
brane is flat. If this is not the case, then there are bound to
be further complications. For instance, one can think of 
two scenarios : (a) an induced, static spherically symmetric
metric on the brane (b) an induced cosmological metric. In both
these cases, the equation for the quantity F would change
drastically through the wave operator defined on the curved
submanifold and also through the intrinsic curvature of the 
induced metric. The general point made in this article about the role
of where we impose initial conditions and the possibility of
focusing along all possible lower dimensional submanifolds
however remains unaltered, though explicit solutions will surely
be very different. 

Finally, it is useful to try and see what can be said about focusing by
looking at the first order equation quoted at the beginning of
the paper. At a general level, if one assumes $M^2 >0$ then we have

\begin{equation}
\nabla^a \theta_a \leq \frac{1}{N-D} \theta_a\theta^a
\end{equation}

 More specifically, for the warped background with a trivially embedded
3--brane (the context we have been discussing here), the first
order equation gives:

\begin{equation}
\left (\partial_0 \theta_0 + \theta_0^2 \right ) = \sum_{\alpha=0}^{3}
\left (\partial_{\alpha}\theta_\alpha +\theta_\alpha^2 \right ) + M^2 e^{2 f(\sigma_0)} 
\end{equation}

The standard analysis for geodesic focusing can be utilised for the
above equation provided we assume that the R.H.S of the above equation
is {\em negative}. If such is the case, then we can say that $\theta_0$
will go to negative infinity within a finite value of the parameter $t$.
We cannot, however, say anything concrete about the behaviour of the
other $\theta_a (a=x,y,z)$ except that they should be such that the
full R. H.S. of the above equation is  negative. 

\section{Concluding remarks}

Let us now conclude with a summary of the results obtained.

$\bullet$ In a warped background, a family of embedded branes with
an initial expansion vector field, can focus along lower
dimensional submanifolds (timelike or spacelike) depending on where
(i.e. on which submanifold) the initial expansions are specified.

$\bullet$ (i) If the initial expansion vector components 
are given on a timelike 2--brane then the surface congruence will have 
a future expansion vector with some (but not all) of its components
diverging to negative infinity. (ii) On the other hand, a specification of
the initial expansion on a two dimensional section of a spacelike 
hypersurface can give rise to a future expansion vector with all
components diverging to negative infinity. (iii) Finally, specifying 
initial conditions on a curve (one dimensional) or a point (zero dimensional)
will lead, once again, to
a future expansion vector with some of its components diverging to
negative infinity. All of the above could be termed as analogs of
the usual geodesic focusing though the situation in (ii) is markedly different
from the ones in (i) and (ii). One is tempted to conjecture (based
on the example discussed) that 
for a $D$ dimensional timelike, embedded hypersurface, all components of the
expansion vector can diverge on a $D-2$ dimensional submanifold.   

$\bullet$ As mentioned before, the above issues change quantitatively 
if we have curvature on the brane or we have a non--trivial embedding
or if we look at Euclidean signature induced metrics. Higher codimension
branes/surfaces involve further complications due to the presence
of more than one normal vector. A detailed treatment of all the above cases 
mentioned is a topic of future investigation and will be communicated in due
course.


\begin{references}
\bibitem{wald} R. M. Wald, {\em General Relativity}, 
Overseas Press (India) Limited (2006)
\bibitem{poisson} E. Poisson {\em A Relativist's toolkit: the mathematics
of black hole mechanics}, Cambridge University Press (2004)
\bibitem{he} S. W. Hawking and G. F. R. Ellis, {\em The Large
Scale Structure of Space-Time}, Cambridge University Press,
Cambridge (1973)
\bibitem{cg} R Capovilla and J. Guven, Phys.Rev. {\bf D52} (1995) 1072-1081 ; E. Zafiris, J.Geom.Phys. {\bf 28} (1998) 271-288;  B. Carter,
Contemp.Math. {\bf 203} (1997) 207-219;
S Kar, 
Phys. Rev.{\bf D54} (1996) 6408-6412 ; {\em ibid.} 
{\bf D53} (1996) 2071-2077 
{\bf D52} (1995) 2036-2043 ; 
S. Kar, {\sf Generalised Raychaudhuri Equations for Strings and Membranes},
Written version of invited talk at IAGRG18 (1996) (IMSc, Chennai Report no 117)
\bibitem{skprd} S. Kar, Phys. Rev. {\bf D55} (1997) 7921-7925
\bibitem{rs} L. Randall and R. Sundrum, Phys. Rev. Lett. {\bf 83}, 3370
(1999); L. Randall and R. Sundrum, Phys. Rev. Lett. {\bf 83}
4690 (1999)
\bibitem{pk06} S. Pal and S. Kar, Class. Quant. Grav. {\bf 23}, 2571-2583 (2006)

\end{references}
\end{document}